\RequirePackage[2020-02-02]{latexrelease}
\documentclass[%
 aip,
rsi,%
 amsmath,amssymb,
preprint,%
]{revtex4-1}

\usepackage{float}
\usepackage{graphicx}
\usepackage{dcolumn}
\usepackage{bm}
\usepackage{color}
\usepackage[version=3]{mhchem} 
\usepackage{subcaption}
\usepackage[utf8]{inputenc}

\begin{document}


\title[]{A method to capture the large relativistic and solvent effects on UV-vis spectra of photo-activated metal complexes}
\author{Joel Creutzberg}
\email{joel.creutzberg@teokem.lu.se}
\affiliation{Division of Theoretical Chemistry, Lund University, Lund, Sweden}
\author{Erik Donovan Hedeg{\aa}rd$^{1,}$}
\email{erdh@sdu.dk}
\affiliation{Department of Physics, Chemistry and Pharmacy, Campusvej 55, 5230 Odense, Denmark}
\date{\today}

\begin{abstract}
\textbf{Abstract:}
We have recently developed a method based on relativistic time-dependent density functional theory (TD-DFT) that allows the calculation of electronic spectroscopy in solution \cite{Creutzberg2022}. This method treats the solvent explicitly with a classical,  polarizable embedding (PE) description. Further, it employs the complex polarization propagator (CPP) formalism which allows calculations on complexes with a dense population of electronic states (such complexes are known to be problematic for conventional TD-DFT). Here we employ this method to investigate both the dynamic and electronic effect of the solvent for the excited  electronic states of  
 \textit{trans}-\textit{trans}-\textit{trans}-\ce{[Pt(N3)2(OH)2(NH3)2]} in aqueous solution. This complex  decomposes into species harmful to cancer cells under light irradiation. Thus,  understanding its' photo-physical properties may lead to a more efficient method to battle cancer. We quantify the effect of the underlying structure and dynamics by classical molecular mechanics simulations, refined with a 
subsequent DFT or semi-empirical optimization on a cluster. Moreover, we  
 quantify the effect of employing different methods to set up the solvated system, e.g., how sensitive the results are to the method used for the refinement, and how large a solvent shell that is required. The electronic solvent effect is always included through a PE potential.  
\end{abstract}
\pacs{Valid PACS appear here}
\maketitle
\section{Introduction}
Platinum (II) complexes are used in cancer treatment since the 1970s. The prototypical complex for cancer treatment is   \textit{cis}-\ce{[Pt(NH3)2Cl2]}, but several complexes with similar motifs are in use today.\cite{rosenberg1969,johnstone2016,imran2018} Unfortunately, the treatments cause severe side-effects.\cite{oun2018} The side effects occur since  \textit{cis}-\ce{[Pt(NH3)2Cl2]} reacts with other bio-molecules in the body than the target molecules inside the cancer cells. To circumvent these side effects it has been suggested to use a so-called  \textit{pro-drug}.\cite{johnstone2016} A pro-drug is inactive (and thus harmless for the body) until activation at the site of the tumor. Octahedral Pt(IV) complexes have been investigated as pro-drugs over the last few years.\cite{johnstone2016,imberti2020,imran2018} These complexes are kinetically stable d$^{6}$ complexes until they are photo-activated. The use of photo-activation combined with a pro-drug has been coined \textit{photo-activated anti-cancer therapy} (PACT).\cite{bonnet2018,imberti2020} 

\begin{figure}[htb!]
\centering
\includegraphics[width=0.3\textwidth]{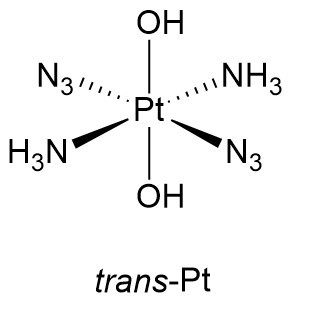}
\caption{Lewis structures of \textit{trans}-\textit{trans}-\textit{trans}-\ce{[Pt(N3)2(OH)2(NH3)2]} (\textit{trans}-Pt)  investigated in this paper.}
\label{fig:lewis}
\end{figure}

Different octahedral Pt(IV) complexes have been investigated for use in PACT. A class of complexes that has received considerable attention is the diazido Pt(IV) complexes.\cite{huayun2019,imberti2020}  One of the most simple of these complexes is shown in Figure \ref{fig:lewis}. The advantage of the diazido complexes is their inertness with respect to bio-reducing agents (such as glutathione)\cite{kratochwil1998}.
The complexes are known to decompose after irradiation, but the mechanistic pathways of this decomposition are still not understood.\cite{imberti2020} A more complete understanding of the photo-physical properties of the Pt(IV) pro-drugs is the next step to further develop PACT. 

Theoretical methods based on time-dependent density functional theory (TD-DFT) can help unravel photo-physical properties.\cite{hedegaard2022} Investigations by Salassa et al.\cite{salassa2009} employed this method to understand and characterize the UV-vis spectrum of  \textit{cis}-\textit{trans}-\textit{cis}-\ce{[Pt(N3)2(OH)2(NH3)2]}: They found that the excitations are mainly of ligand-to-metal-charge-transfer (LMCT) character, involving orbitals on azide ligands and platinum. The resulting excited states were found to be dissociative with respect to \ce{N3-}. Triplet states were also speculated to be involved in the decomposition. 
In a later paper by Sokolov et al.\cite{sokolov2011} similar results were obtained  both for \textit{trans}-\textit{trans}-\textit{trans}-\ce{[Pt(N3)2(OH)2(NH3)2]} and \textit{cis}-\textit{trans}-\textit{cis}-\ce{[Pt(N3)2(OH)2(NH3)2]}, albeit with the difference that the excitation facilitated the release of two azide radical ligands. 

Calculations of the excited state manifold of platinum complexes are generally expected to require relativistic effects. These effects have mostly been included with either scalar relativistic Douglas-Kroll-Hess (DKH) to second order\cite{sokolov2011} or effective core potentials (ECPs).\cite{salassa2009} None of these methods include spin-orbit coupling (SOC) explicitly. In a recent investigation, we investigated the effect of  SOC on calculated UV-vis spectra for the Pt(IV) complexes \textit{trans}-\textit{trans}-\textit{trans}-\ce{[Pt(N3)2(OH)2(NH3)2]} and \textit{cis}-\textit{trans}-\textit{cis}-\ce{[Pt(N3)2(OH)2(NH3)2]}, employing a four-component linear response framework.\cite{creutzberg2020}  One of our main findings was that the SOC  led to a much denser manifold of electronic states and many of these states were of mixed singlet-triplet character.  Similar findings were also obtained in a recent study by Freitag and González, using a relativistic DMRG-SCF method to study the excited state dissociation reactions.\cite{freitag2021}   

Thus, explicit inclusion of SOC is important to correctly access the excited state manifold of the Pt(IV) complexes used in PACT. Yet, even with inclusion of SOC, the solvent (usually water) employed in  experimental work may also have a large impact on the excited states and  UV-vis spectra. Accordingly, the next step is to investigate the effect of the solvent: Although we will not study the full dissociation mechanism here, we note that there are strong indications that the decomposition mechanisms of the Pt(IV) complexes are solvent dependent.\cite{phillips2009,ronconi2008,ronconi2011a} However,  previous theoretical studies either ignored\cite{sokolov2011,creutzberg2020,freitag2021} the solvent or included it through continuum models\cite{salassa2009}, which are known to be inaccurate for solvents prone to form hydrogen bonds. 

Over the last years, we and others have developed models that explicitly include the solvent, employing a polarizable embedding (PE) model.\cite{olsen2010a,sneskov2011a,hedegaard2013a,hedegaard2015a,hedegaard2016b,steinmann2018}
The PE model includes the environment classically with multipoles and polarizabilities, allowing mutual polarization between the quantum mechanical and classical subsystems.  Within non-relativistic frameworks, the PE models have been shown to work well for electronic spectroscopy\cite{bondanza2020,slipchenko2010,sneskov2011b,defusco2011,beerepoot2014}. The PE models were only recently introduced to a relativistic framework
\cite{krause2016,hedegaard2017,Creutzberg2022} and applications of this methodology have therefore been rare.  Initial calculations  have demonstrated that the electronic solvent effect on \textit{trans}-\textit{trans}-\textit{trans}-\ce{[Pt(N3)2(OH)2(NH3)2]} is substantial.\cite{Creutzberg2022} However,  structural and/or dynamical effects of the solvent have not been investigated. In this investigation, we undertake a more systematic investigation of solvent effects for the  \textit{trans}-\textit{trans}-\textit{trans}-\ce{[Pt(N3)2(OH)2(NH3)2]} complex.    We will mainly address how the structural changes in the environment influence the spectra by performing molecular dynamics simulations. The dynamics are obtained through classical simulations, but since no force field exists for the platinum complexes, we have constructed one based on calculated charges and the molecular Hessian. We have set up a method in which snapshots extracted from the classical simulation are refined by cluster calculations, employing first a semi-empirical method and then DFT. Part of this paper is also  devoted to quantifying to what degree the calculated UV-vis spectra depend on the QM methods employed in this setup.  Finally, we also investigate the effect of including water molecules explicitly in the QM region. 
 
\section{Theory}

In the PE model, the system is divided into a region that is treated using methods from quantum chemistry (QM), and a region treated classically. The classical region is parameterized by electrostatic multipoles and point-polarizabilties, calculated from a quantum mechanical method by fragmenting the solvent into individual molecules. The total energy expression is then 
\begin{align}
E = E_{\mathrm{QM}} + E_{\mathrm{es}} + E_{\mathrm{ind}} + E_{\mathrm{env}} , 
\label{eq:total_energy}
\end{align}
where the first term is the QM energy expression, given as
 \begin{align}
 E_{\mathrm{QM}} = \sum_{pq}h_{pq}D_{pq} + \sum_{pq}j_{pq} D_{pq} + E_{\mathrm{xc}}[\rho] + E_{\mathrm{nn}}.   
\label{eq:energy_func}
\end{align}
We have in Eq.~\eqref{eq:energy_func} used a second quantization formalism. Thus, $D_{pq}$ is an element of the one-electron reduced density matrix, $h_{pq}$ is an integral over the kinetic energy and nuclear attraction one-electron operators, and  $j_{pq}$ contains the Coulomb integral for electron repulsion (and scaled exchange integrals if hybrid functionals are used). Finally, $E_{\mathrm{xc}}[\rho]$ and $E_{\mathrm{nn}}$  contain the correlation--exchange functional and nuclear repulsion, respectively. For brevity, the former is only a functional of the density, $\rho$. We refer to  the literature for explicit expressions\cite{saue2002} of the terms involved in Eq.~\eqref{eq:energy_func}. Note that in the second-quantization formalism, non-relativistic, two- or four-component frameworks are \textit{formally} equivalent and the form of Eq.~\eqref{eq:energy_func} is therefore similar in all three cases. However, the form of the integrals will differ between non-relativistic, two- or four-component frameworks and modifications of the $E_{\mathrm{xc}}[\rho] $ term are also required in a relativistic framework.  The largest difference is seen in the one-electron integrals, $h_{pq}$, where the integrals in a four-component framework are over the operator 
\begin{align}
	\hat{h} = \hat{h}_{D} + \hat{V}_{\mathrm{ext}} =  
	\left(\begin{array}{cc} \bm{0}_2 & c(\bm{\sigma}\cdot \hat{\mathbf{p}}) \\
		c(\bm{\sigma}\cdot \hat{\mathbf{p}})   &   - 2c^{2} \mathbf{I}_2
	\end{array}\right) + 
	\left(\begin{array}{cc}  V_{\mathrm{ext}}\mathbf{I}_2  & \bm{0}_2 \\
		\bm{0}_2   &  V_{\mathrm{ext}}\mathbf{I}_2
	\end{array}\right) . 
	\label{eq:h_D}
\end{align}
Here, $\bm{\sigma}$ denotes the Pauli spin  matrices, $\mathbf{I}_2$ and  $\mathbf{0}_2$ are $2\times 2$ unit and zero matrices while $c$ is the speed of light. Finally,  $V_{\mathrm{ext}}$ is an external potential that here is reduced to the nuclei-electron attraction (see Refs.~\cite{hedegaard2017} and \cite{Creutzberg2022} for further details). As can be inferred from Eq.~\eqref{eq:h_D}, the dimension of $\hat{h}$ is $4\times 4$ and hence the wave function will also have four components. These four components are usually divided into a large and small component wave function, each with two components. In this paper, we only employ a four-component wave function to a very limited extent and we instead focus on the two-component eXact decoupling (X2C) method implemented in the DIRAC program\cite{saue2020}.  The decoupling method uses the matrix transformation 
\begin{align}
	\label{x2c-U}
	\mathbf{U}^{\dagger}\mathbf{h}_D\mathbf{U} = \left( \begin{array}{cc} \mathbf{h}_{++} & 0  \\
		0 & \mathbf{h}_{--}   \end{array} \right) , 
\end{align}	
to decouple the large and small component wave functions\cite{saue2011}. The decoupling in  Eq.~\eqref{x2c-U} allows us to focus only on the positive energy solutions,  
which reduces the computational cost significantly.

The energy $E_{\mathrm{es}}$ in Eq.~\eqref{eq:total_energy} accounts for the interaction between the electrostatic multipoles in the environment and the QM  density, whereas the energy, $E_{\mathrm{ind}}$, accounts for mutual polarization of the QM and MM densities (the latter through the point polarizabilities). These two energies can be represented by the operators, $\hat{V}^{\mathrm{es}}$, and $\hat{V}^{\mathrm{ind}}$ that are added to the vacuum Kohn-Sham operator 
 \begin{align}
	\hat{f}^{\mathrm{tot}} & = \hat{f}_0 + \hat{V}^{\mathrm{es}} +  \hat{V}^{\mathrm{ind}} =  \hat{f}_0 + \hat{V}^{\mathrm{es}} - \bm{\mu}^{\mathrm{ind}} \hat{\bm{\mathcal{E}}}^{\mathrm{e}}   \label{eq:fock_pe-1}.
\end{align}
In this formalism, $\bm{\mu}^{\mathrm{ind}}$ is a vector containing the induced dipole moments calculated as\cite{applequist1972} $ \bm{\mu}^{\mathrm{ind}} = \mathbf{R}^{Relay}\bm{\mathcal{E}}$. The definition of $\mathbf{R}^{Relay}$ can be found in the literature (see e.g.~Ref \citenum{hedegaard2017}) whereas $\bm{\mathcal{E}}$ is the total electric field, here given for a specific site $s$ in the environment
\begin{align}
	\label{eq:total_field}
	\hat{\bm{\mathcal{E}}}_s = \hat{\bm{\mathcal{E}}^{\mathrm{e}}}_s + \bm{\mathcal{E}}^{\mathrm{nuc}}_s +  \bm{\mathcal{E}}^{\mathrm{es}}_s  .
\end{align} 
The total electric field has components from nuclei in the QM region, $\hat{\bm{\mathcal{E}}}^{\mathrm{n}}$, electrons in the QM region, $\hat{\bm{\mathcal{E}}}^{\mathrm{e}}$, and the multipoles on the remaining sites, $\hat{\bm{\mathcal{E}}}^{\mathrm{es}}$, where  we again refer to the literature for details \cite{hedegard2013}.  Below, we will use the short-hand form $\hat{V}^{\text{ind}}$ for the term involving the induced dipoles. 
   
In this work we employ linear response theory\cite{olsen1985,helgaker2012} to calculate UV-vis spectra. The linear response formalism is often denoted TD-DFT. In this formalism, we solve the equation 
\begin{align}
	\label{eq:solution-vector}
\bm{\kappa}^{z} = - \Bigl(\mathbf{E}^{[2]} - z\mathbf{S}^{[2]}\Bigr)^{-1} \mathbf{E}^{[1]}_X .
\end{align} 
$\mathbf{E}^{[1]}_X$ is a property gradient (here employing the dipole operator), $\mathbf{S}^{[2]}$ the metric, $\bm{\kappa}^{z}$ the solution vector, and $\mathbf{E}^{[2]}$ is the electronic Hessian, defined as    
\begin{align}
    \mathbf{E}_0 ^{[2]} = 
\begin{pmatrix}
    \mathbf{A} & \mathbf{B}       \\
     ~\mathbf{B}^* &  ~\mathbf{A}^*              
\end{pmatrix} . 
\end{align}
Expressions for the \textbf{A} and \textbf{B} terms can be found elsewhere: the vacuum forms, here denoted  $\textbf{A}^{\text{vac}}$ and $\textbf{B}^{\text{vac}}$, are provided by Salek et al.\cite{salek2005} We define the additional terms due to a PE environment below.  Since regular linear response theory is known to suffer from problems in regions with a a high density of states (which is common on relativistic calculations\cite{villaume2010,Creutzberg2022}), we will generally employ the complex polarization propagator (CPP).\cite{norman2001,kristensen2009,villaume2010,kauczor2014} For the CPP, we set $z = \omega + i\gamma$, where $\gamma$ is a phenomenological (inverse) lifetime; regular linear response theory corresponds to setting $\gamma = 0$, i.e., $z = \omega$.  With  $z = \omega + i\gamma$ we have access to the imaginary part of the frequency dependent polarizability tensor, $\mathrm{Im}[\bm{\alpha}]$, which is directly related to the cross-section $\sigma (\omega)$ of the signal in electronic spectroscopies 
\begin{equation}
	\label{eq:cross_section}
	\sigma (\omega)= \frac{4\pi \omega}{3c}\text{Im}[\alpha_{xx}+\alpha_{yy}+\alpha_{zz}] . 
\end{equation}
The calculation of the signal cross section is done with an input-defined frequency, $\omega$, meaning that the CPP method has the same cost in all frequency ranges. This is unlike traditional response solvers, which obtain the frequency by solving for a number of roots, starting from the lowest ones (thus the total number of roots  may become very high for regions with dense population of electronic states).   

We have previously shown how the modification of Eq. \ref{eq:solution-vector} due to the PE framework leads to  modified linear response equations for both regular linear response\cite{olsen2010a,hedegaard2017} and the CPP model.\cite{pedersen2014,Creutzberg2022} In short, these changes are, 
\begin{align}
A_{ai;bj} & = A^{vac}_{ai;bj} + \langle  0 \vert [\hat{q}^{\dagger}_{ai},[ \hat{q}_{bj}, \hat{V}^{\mathrm{es}} + \hat{V}^{\mathrm{ind}}]] \vert 0 \rangle + \langle  0 \vert [\hat{q}^{\dagger}_{ai}, \tilde{V}^{\mathrm{ind}}] \vert 0 \rangle \\
B_{ai;bj} & =	B^{vac}_{ai;bj} +  \langle  0 \vert [\hat{q}_{ai},[ \hat{q}_{bj}, \hat{V}^{\mathrm{es}} + \hat{V}^{\mathrm{ind}}]] \vert 0 \rangle + \langle  0 \vert [\hat{q}_{ai}, \tilde{V}^{\mathrm{ind}}] \vert 0 \rangle ,  \label{response-E}  
\end{align}
where $\hat{V}^{\text{es}}$ and $V^{\text{ind}}$ were defined in Eq.~\eqref{eq:fock_pe-1} and  $\tilde{V}^{\mathrm{ind}}$ is a transformed form of $V^{\text{ind}}$ (a more detailed derivation can be found in the literature). \cite{saue2003,salek2005,Creutzberg2022,hedegaard2017}  The first term accounts for the effect due to the multipoles and the ground-state polarization. The second term accounts for the change in solvent response due to the change of electron density in the QM system during an electronic excitation. 

Finally, we introduce effective external field effects (EEF)\cite{jensen2005,list2016}.  This  contribution arises as the external magnetic field that induces an electronic transition, also induces a field, $\tilde{\bm{\mathcal{E}}}(t)$, within the environment. This can be accounted by modifying  the total induced field in Eq.~\eqref{eq:total_field} to $\tilde{\bm{\mathcal{E}}}^{\mathrm{tot}}_s = \tilde{\bm{\mathcal{E}}}^{\mathrm{e}}_s + \bm{\mathcal{E}}^{\mathrm{nuc}}_s +  \bm{\mathcal{E}}^{\mathrm{es}}_s + \tilde{\bm{\mathcal{E}}}_{s}(t) $ within the linear response framework.   The result is that the property gradients are modified (see e.g.~Ref.~\citenum{hedegaard2017} for a derivation)   
\begin{align}
	\bar{E}^{[1]}_{ai,X} = \langle 0 \vert [\hat{q}_{ai},\hat{X} ] \vert 0 \rangle + \frac{\text{d}\bm{\mathbf{\mu}}^{\mathrm{ind}}_{\text{ext},X}(\omega_k)}{\text{d}\bm{\mathcal{E}}_X (\omega_k)}  \langle 0 \vert [\hat{q}_{ai},\hat{\bm{\mathcal{E}}}^{\mathrm{e}} ] \vert 0 \rangle .  
\end{align}
The last term, containing $\bm{\mu}^{\mathrm{ind}}_{\text{ext},X}(\omega_k) = \mathbf{R}^{Relay}\bm{\mathcal{E}}_X (\omega_k)$, is responsible for  the EEF effect.

\section{Computational details}

\begin{figure}[htb!]
\includegraphics[width=0.45\textwidth]{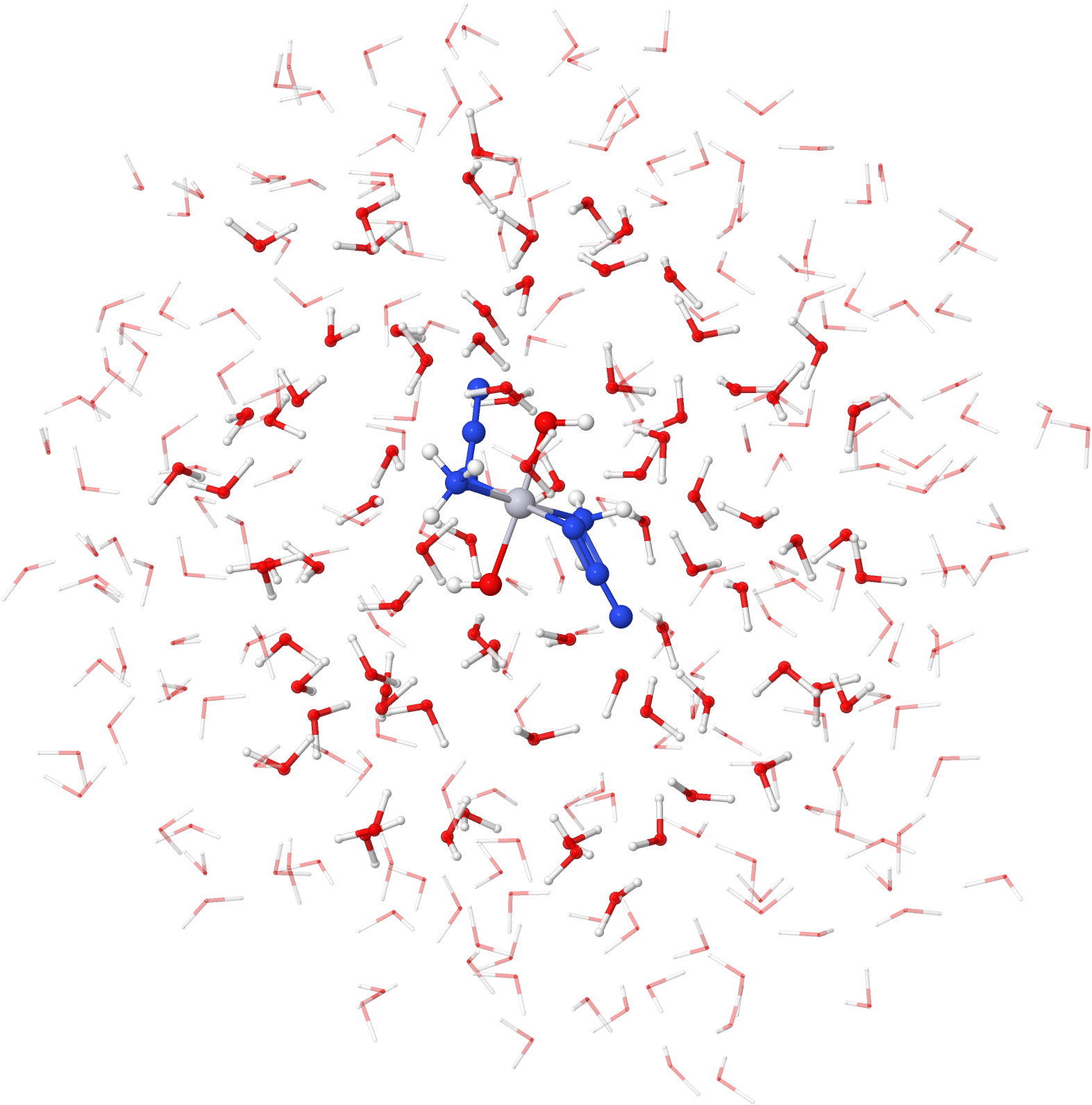}
\includegraphics[width=0.45\textwidth]{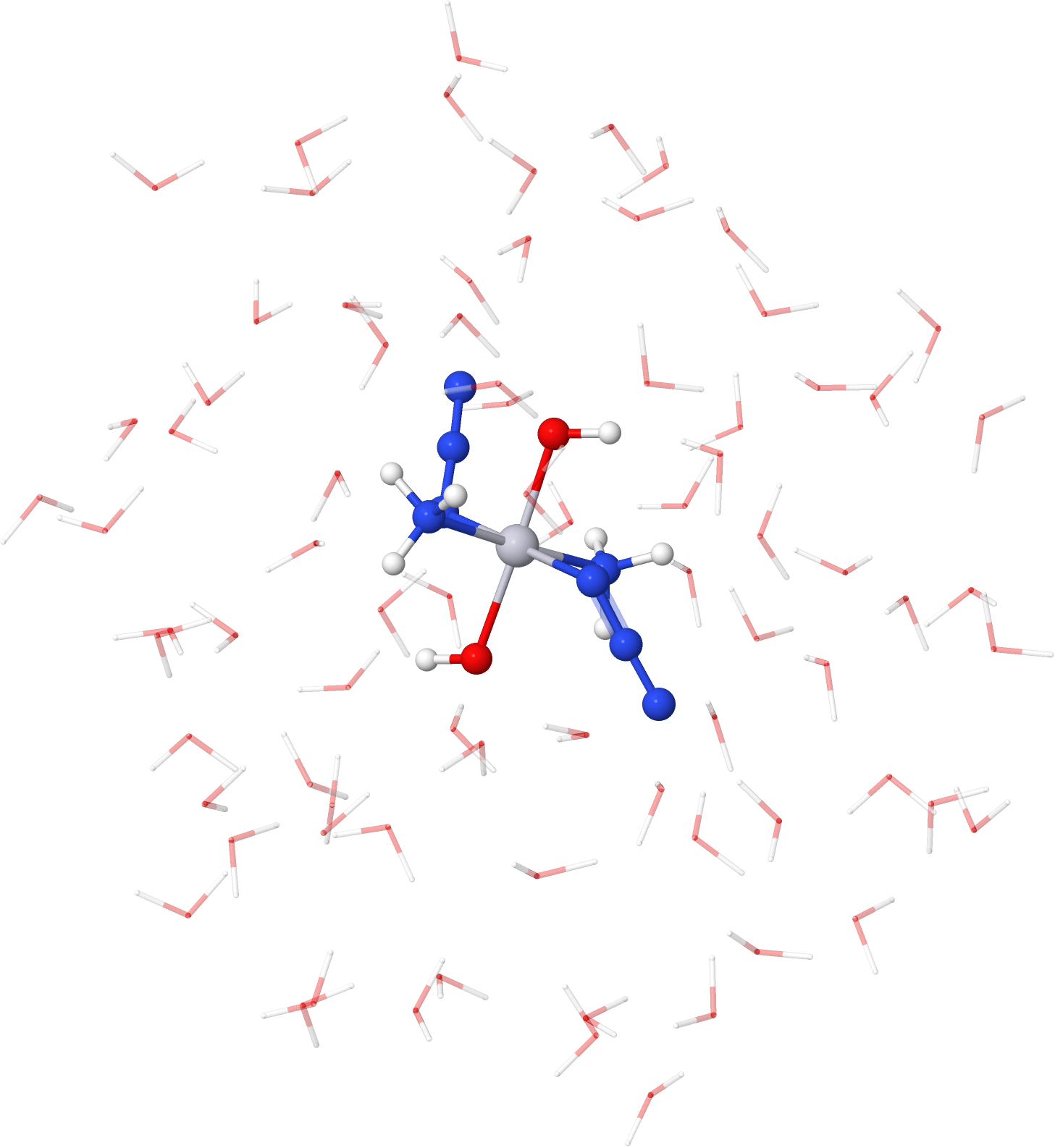}
\caption{Optimized structure of \textit{trans}-Pt complex, in water, employing PBEh-3c.}
\label{fig:pt_in_water}
\end{figure}

\textbf{Static structures and potentials of \textit{trans}-Pt complex in water:} Since the structures obtained during the MD simulations are purely classical, we refine these structures through quantum mechanical optimizations on smaller clusters (more details are given in the next subsection). We initially employed a fast semi-empirical method for this refinement to save computer time. The structures were subsequently refined with DFT. We decided to employ a single structure to investigate  the effect  of the additional DFT refinement on the calculated UV-vis spectrum, before embarking on optimizing a large number of snapshots. Rather than employing a random snapshot from the classical MD, the structure was taken from our previously optimized solvated \textit{trans}-\textit{trans}-\textit{trans}-[\ce{Pt(N3)2(OH)2(NH3)2]} (\textit{trans}-Pt) complex\cite{Creutzberg2022}. We recapitulate the optimization here: we employed QM/MM in a box (35 $\times$ 35 $\times$ 35 \AA) of explicit (frozen) water molecules, optimizing only the \textit{trans}-\textit{trans}-\textit{trans}-[\ce{Pt(N3)2(OH)2(NH3)2]} complex. This optimization was  followed by two subsequent optimizations. First, we made a cutout including the water molecules within 6 {\AA} from the platinum complex. Structure optimizations of this cluster were then carried out relaxing the platinum complex and all water molecules within 4 {\AA}, while the rest were kept frozen. This was done with the PBEh-3c \cite{grimme2015} method. Next, we refined this optimization  with the BP86 functional\cite{Becke1988} and a  def2-sv(p)\cite{weigend2005,weigend2006} basis set. We chose this functional since it is computationally efficient and has been shown to provide structures similar to B3LYP in accuracy for transition metal complexes.\cite{buhl2008} These calculations were done in ORCA.\cite{orca}  Finally, the 6  {\AA}  sphere was re-inserted into the original box and a 10 {\AA}   sphere was cut out (see Figure \ref{fig:pt_in_water} were the PBEh-3c systems are shown as examples). The largest cluster of 10 {\AA} contains 282 water molecules, whereas the 6 {\AA} cluster contains 79 water molecules. Since we also employed these structures in Ref.~\cite{Creutzberg2022}, we can re-use previously obtained PE potentials.\cite{Creutzberg2022}

\textbf{Force field for \textit{trans}-Pt:}  We developed a force field (bonded model) for \textit{trans}-pt complex, employing the MCPB.py program\cite{mcpb}. This program extracts bonding parameters from the molecular Hessian and employs the RESP model for the electrostatics.   The structure optimization, force constants, and Merz-Kollman RESP charges were obtained from a calculation with the B3LYP\cite{Becke1988,Lee1988,Becke1993} functional in Gaussian 16 \cite{g16}. The 6-31G* \cite{petersson1988,petersson1991} basis set was employed for the ligands while the effective core potential SDD\cite{Andrae1990a} was employed for the Platinum atom. Default radii were used for all atoms. A final adjustment of the force field parameters was done by setting the angle between the nitrogen atoms in the N$_3^-$ ligands to 180 degrees, in order to ensure that they were kept linear during the simulations. 

\textbf{Dynamics and polarizable embedding potentials:} Using the constructed force field the complex was solvated in a box (10 $\times$ 10 $\times$ 10 \AA) of explicit TIP3P water molecules. A minimization was then done preceded by two MDs to ensure the platinum complex maintained a reasonable structure: first a MD for 20 fs at constant pressure  with SHAKE.\cite{ryckaert1977} After ensuring the integrity of the platinum complex, we continued this MD for 1 ns. Thereafter an equilibration simulation was done at constant pressure for 2 ns before the final production MD was done for 20 ns (also at constant pressure). All calculations were done in AMBER 16 \cite{amber16}. From the trajectory, 49 structures were extracted with a minimum time separation of 0.2 ns. After this, the same QM optimization procedure as for the first single structure (\textit{trans}-Pt complex in water)  was carried out on 6 {\AA}  sphere cutouts from the structures, but no re-insertion into the original box was done. Thus, the final systems are all 6 {\AA}, which (as we show below) is sufficient. Selected structures as well as the force-field parameters are provided as a zenodo repository.\cite{joel_creutzberg_2023_7791336}

Embedding potentials were calculated for the snapshots described above, using the B3LYP functional with  the A-6-31PGP basis set and included multipole moments up to quadropoles and anisotropic polarizabilities. Fragmentation of the environment was carried out using the PyFrame \cite{pyframe:0.4.0} script, where each fragment in the environment consisted of a water molecule.  The localized multipoles and polarizabilities were obtained from the LoProp method\cite{gagliardi2004} as implemented in DALTON \cite{olav_vahtras_2014_13276,olsen2020} (the level of theory employed for the potentials is identical to the potentials from Ref.~\cite{Creutzberg2022}). 

\textbf{Frequency dependent polarizablities}
CPP calculations were carried out in DIRAC \cite{DIRAC22} in the UV-Vis range (6--2 eV) using the CAM-B3LYP\cite{takeshi2004} functional and a $\gamma$ value of 1000 cm$^{-1}$ (0.124 eV). We  employed the following Hamiltonians: Levy-Leblond (non-relativistic limit), Dyall's spinfree Hamiltonian (scalar relativistic) and X2C. We initially also carried out linear response calculations with both four-component Dirac-Kohn-Sham and X2C Hamiltonians employing 80 roots with a conventional linear response solver. This was done in order to ensure that X2C was sufficient to reproduce four-component results. The calculated four-component spectra are close to identical to the X2C spectra. Seeing that larger errors are expected from the choice of functional or setup of the environment, we have not included the four-component calculations in the paper (they are provided in the supporting information). 
Polarizable embedding (PE) was included in the calculations using the {\sc PeLib} library (which is interfaced to DIRAC), moreover, effective external field calculations (EEF) were included. All calculations employed the dyall.v2z \cite{dyall2010} basis set for the platinum atom, while the ligands employed a def2-sv(p) basis set.  For a few snapshots, we additionally carried CPP calculations with B3LYP (and the same double-zeta basis sets as above) or with CAM-B3LYP and triple zeta basis sets (the results from these calculations are shown in Figures S4 and S5).   The basis sets were always used uncontracted.

We will in the following not discuss the non-relativistic results, since they (as expected) were quite different from the X2C calculations.

\section{Results and discussion}  
We first discuss the effect of optimizing the Pt(IV) complex in a small cluster of water molecules after the initial solvation.  We next compare the spectra (and solvent shifts) resulting from employing the X2C and scalar relativistic Hamiltonians for PE regions of different sizes. We additionally investigate the effects of extending the QM region with selected water molecules. These initial investigations will serve to justify the method used for the dynamics, which is done in the last subsection.    

\subsection{Structural solvent effect from method employed in refinement.}
\begin{figure}[hbt!]
\hspace{-2cm}
 \begin{subfigure}[b]{0.45\textwidth}
\includegraphics[width=1.3\textwidth]{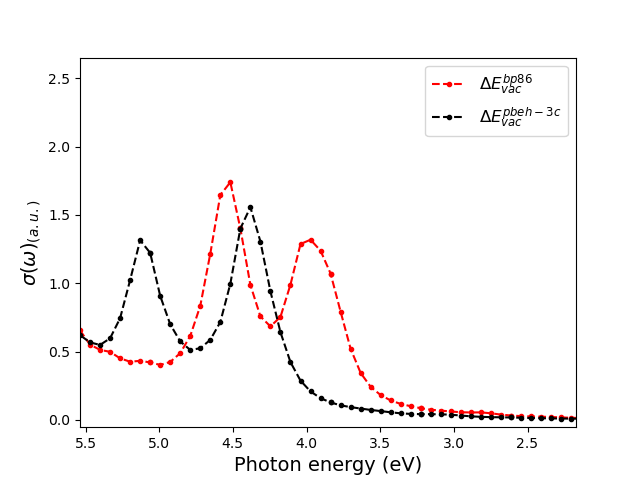}
\caption{\footnotesize}
\end{subfigure}
~~~~~~~~ \begin{subfigure}[b]{0.45\textwidth}
\includegraphics[width=1.3\textwidth]{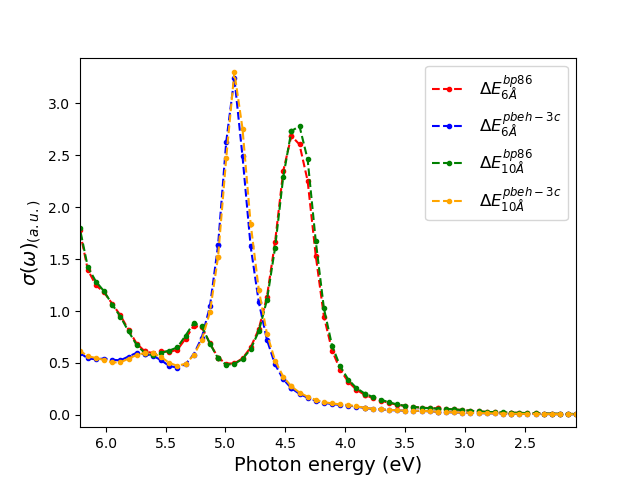}
\caption{\footnotesize }
\end{subfigure}
 \caption{\footnotesize (a) UV-vis absorption spectra calculated in vacuum with  structures obtained from PBEh-3c and BP86 (in solvent). (b) UV-vis absorption spectra calculated with the structures from (a) including PE for 6 and 10 \AA. All spectra are calculated with  X2C-CAM-B3LYP.}
\label{fig:bp86_pbe3h_x2c}
\end{figure}

We start by comparing the spectra obtained from a PBEh-3c optimized structure to those optimized with BP86. To properly separate structural and electronic solvent effects, we first compare the two spectra without a PE description of the environment, i.e., a vacuum calculation on the two structures (Figure \ref{fig:bp86_pbe3h_x2c}a).  Since we previously showed that the electronic effect due to PE can significantly change the spectrum, we repeated the comparison while including a PE environment within 6 and 10 {\AA}  spheres  (Figure \ref{fig:bp86_pbe3h_x2c}b). Yet, both vacuum and solvated UV-vis spectra show that the spectra obtained for the BP86 optimized structure are significantly red-shifted in comparison to the PBEh-3c spectra (peak positions for the individual calculations are given in Table \ref{tab:bp86_pbe3h}): in the vacuum case, the red-shifts are 0.61 and 0.41 eV, respectively.  When including the electronic solvent effect with PE, the corresponding shifts are both  0.48 eV for the 6 {\AA} system, whereas they are 0.41 and 0.54 eV for the 10 {\AA} system. The absorption cross sections also change significantly from vacuum to solvent and this is discussed further in next subsection. Here we only note that for the absorption cross sections in  vacuum,  the first peak is 0.24 a.u. higher for the PBEh-3c structure compared to the first peak of the BP86 structure, while the second peak of PBEh-3c is 0.42 a.u. lower than that of BP86 (see Figure \ref{fig:bp86_pbe3h_x2c}a and Table \ref{tab:bp86_pbe3h}). When including the 6 Å environment, similar changes are observed, with the first peak from the PBEh-3c structure being  0.56 a.u. higher and the second peak 0.26 a.u. lower when comparing to the peaks of the spectra from the BP86 structure (Figure \ref{fig:bp86_pbe3h_x2c}b and Table \ref{tab:bp86_pbe3h}). For the 10 Å environment, the PBEh-3c structure again has a higher absorption cross section for the first peak (0.52 a.u.), and lower (0.29 a.u.) for the second peak compared to BP86 (Figure \ref{fig:bp86_pbe3h_x2c}b and Table \ref{tab:bp86_pbe3h}).

\begin{figure}[hbt!]
\includegraphics[width=0.50\textwidth]{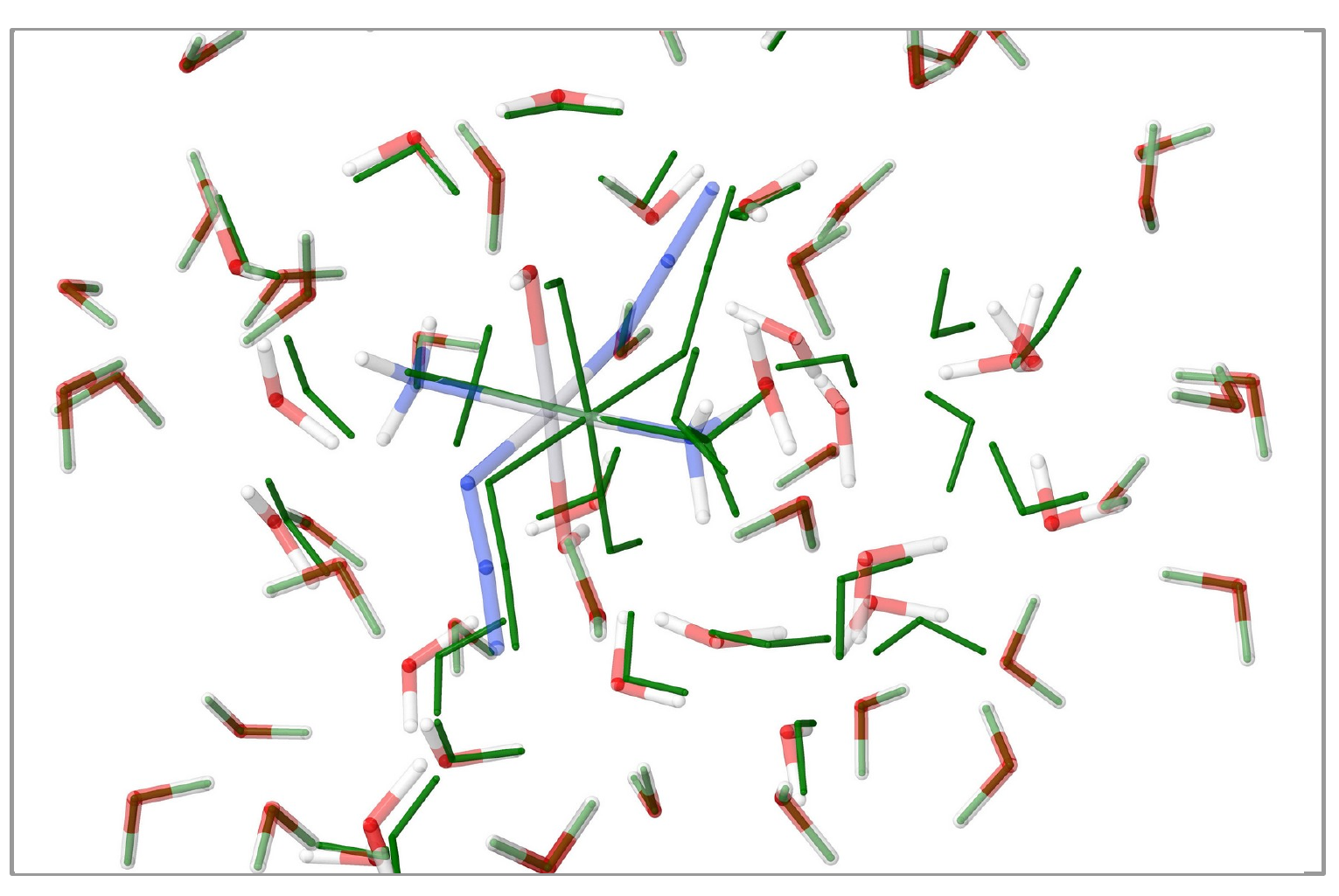}
\caption{\footnotesize Solvated \textit{trans}-Pt system optimized with PBEh-3c (green) or BP86. Only parts of the full 6 {\AA} systems are shown. \label{fig:bp86_pbe3h_struct}}
\end{figure}

In conclusion, both the underlying structure and the electronic contribution of the solvent have large effects on the resulting UV-vis spectra. These large effects  prompted us to investigate the underlying structures. These are shown as overlays in Figure \ref{fig:bp86_pbe3h_struct}. In this figure, some of the water molecules can be seen to overlap exactly and these corresponds to the ones kept frozen in the optimizations. However, both water molecules close to the \textit{trans}-Pt complex and the complex itself show significant structural changes after optimization with BP86.  One of the goals of employing a semi-empirical method for the refinement was to investigate whether the snapshots obtained from the classical dynamics can be refined in a computationally cheap manner.   
Unfortunately, the changes in the structures combined with accompanying  shifts in both energies and cross sections strongly indicate that this is not the case. Thus, we here employ the higher level of theory, e.g.,  BP86. We note for the particular structure employed here, the highest peak of the  BP86 result in Figure \ref{fig:bp86_pbe3h_x2c}b is only 0.03 eV off from the experimental peak at 4.35 eV\cite{mackay2006}, while that of PBEh-3c predicts a peak that is 0.58 eV (33 nm) shifted from the experiment. One of the peaks in the vacuum spectrum of PBEh-3c is also at 4.38 eV, but seeing that the results shift to 4.93 eV due to the electronic effect of the solvent, this result is likely fortitous. In fact, we  show in a section below, that the result from BP86 (including PE) also is likely to be fortuitous, since including dynamical effects will move the result slightly away from the experimental maximum. 

\begin{table}[htb!]
	\centering
	\caption{Selected peak maxima positions $\Delta E$  (in eV)  for the solvent optimized complex with PBEh-3c and BP86 (the corresponding spectra are given in Figure \ref{fig:bp86_pbe3h_x2c}). The absorption cross sections, $\sigma(\omega)$, are given in parantheses.  \label{tab:bp86_pbe3h}}
	\begin{tabular}{llcc}
		\hline
		\hline  \\[-2.0ex]
		Method & ~~~~Environment & ~~~~$\Delta E_1$ [$\sigma(\omega)$]~~~~ & $\Delta E_2$ [$\sigma(\omega)$]  \\ 
		\hline \\[-1.5ex]
		PBEh-3c & ~~~~vac             &  ~~~~4.38~~(1.557)~~~~  &  5.13~~(1.315)            \\[0.5ex]
		BP86   & ~~~~vac             &  ~~~~3.97~~(1.319)~~~~  &  4.52~~(1.740)            \\[0.5ex]
		PBEh-3c & ~~~~solv.~(6 \AA)   &  ~~~~4.93~~(3.243)~~~~  &  5.74~~(0.599)            \\[0.5ex]
		BP86   & ~~~~solv.~(6 \AA)   &  ~~~~4.45~~(2.685)~~~~  &  5.27~~(0.858) \\[0.5ex]
		PBEh-3c & ~~~~solv.~(10 \AA)  &  ~~~~4.93~~(3.297)~~~~  & 5.67~~(0.597)            \\[0.5ex]
		BP86   & ~~~~solv.~(10 \AA)  & ~~~~4.38~~(2.779)~~~~  &  5.27~~(0.885) \\[0.5ex]   
		\hline 
	\end{tabular}
\end{table}

\subsection{Solvent shifts with different choices of Hamiltonian} 
\begin{figure}[hbt!]
\hspace{-2cm}
 \begin{subfigure}[b]{0.45\textwidth}
\includegraphics[width=1.25\textwidth]{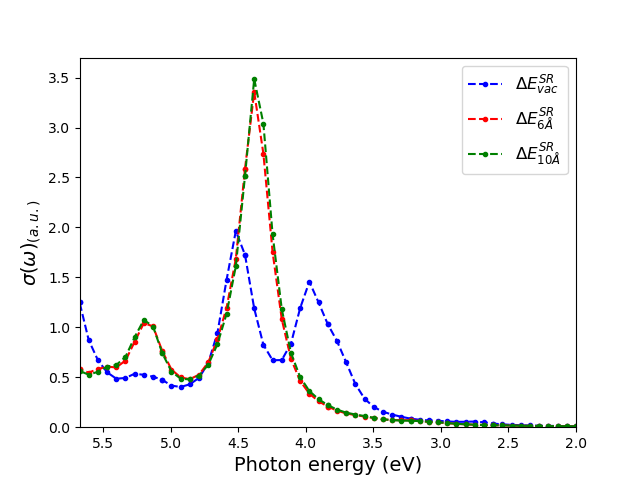}
\caption{\footnotesize }
\end{subfigure}
 ~~~~~~~~~\begin{subfigure}[b]{0.45\textwidth}
\includegraphics[width=1.25\textwidth]{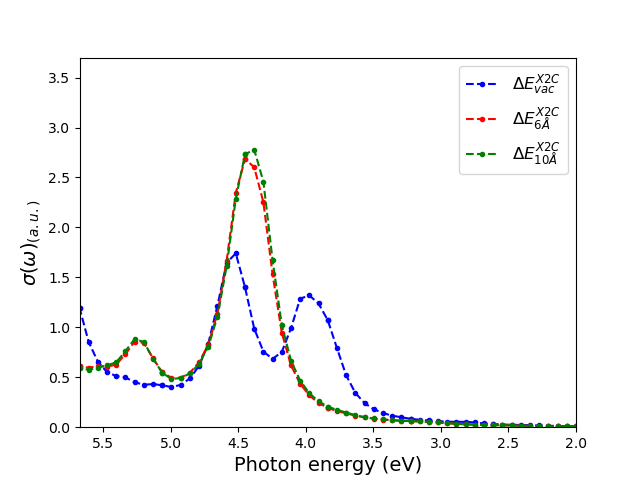}
\caption{\footnotesize }
\end{subfigure}
 \caption{\footnotesize UV-vis spectra calculated (a) scalar relativistic and (b) X2C Hamiltonians, comparing the solvation effects for different levels of environments (from vacuum to a sphere  of 10 {\AA}, all taken from the BP86 optimized structure). All calculations were done X2C-CAM-B3LYP and with structures obtained from BP86 (in solvent).}
\label{fig:solvent_effects}
\end{figure}

In the next series of calculations, we compared the environmental effect obtained by different choices of the underlying Hamiltonian: we either included only scalar relativistic effects (SR) or employ X2C. Systems with both 6 {\AA} and 10 {\AA} solvation spheres are investigated (but we exclusively use the structure refined with BP86). The resulting UV-vis spectra are shown in  Figure \ref{fig:solvent_effects} (a)--(b) and the peak positions are provided in Table \ref{tab:solvated_static_struct_bp86}. From vacuum to solvent, it is seen that the inclusion of PE yields a significant change in the spectra, but this change is qualitatively similar for SR and X2C Hamiltonians: the two distinct peaks in the vacuum spectra around 3.5--4.5 eV becomes a single peak, whereas a new peak appears at higher energies in the solvated spectra. Thus, we can conclude that the difference is not related to the inclusion of SOC, but exclusively an effect of the solvent. To understand the observed difference, we analyzed the transitions in the response vectors, employing the 10 {\AA} system. All peaks are LMCT transitions involving donor orbitals on the \ce{N3-} ligands and acceptor orbitals on the Pt center. The $d$-orbitals contributing to both peaks in the X2C spectra are along the Pt-\ce{N3-} bond axis. Meanwhile, the vacuum spectra have contributions from the $d$-orbital along Pt-\ce{OH-} axis and a $d$-orbital in the \ce{N3-}-\ce{NH3}  plane.  The difference between the contributing $d$-orbitals can explain the differences between vacuum and PE spectra (the SR and X2C calculations employ the same structure for the \textit{trans}-Pt complex). 
\begin{table}[htb!]
	\centering
	\caption{Selected peak maxima positions $\Delta E$  (in eV)  for the solvent optimized complex with PBEh-3c and BP86 (the corresponding spectra are given in Figure \ref{fig:solvent_effects}). The absorption cross sections, $\sigma(\omega)$, are given in parantheses.  \label{tab:solvated_static_struct_bp86}}
	\begin{tabular}{llcc}
		\hline
		\hline  \\[-4.0ex]
		Method & ~~~~Environment & ~~~~$\Delta E_1$ [$\sigma(\omega)$]~~~~ & $\Delta E_2$ [$\sigma(\omega)$]  \\ 
		\hline \\[-3.5ex]
		SR   & ~~~~vac                  &  ~~~~3.97~~(1.456)~~~~   &  4.52~~(1.964)    \\[0.5ex]
		SR   & ~~~~solv.~(6 \AA)        &  ~~~~4.38~~(3.356)~~~~   &  5.20~~(1.043)    \\[0.5ex]
	    SR   & ~~~~solv.~(10 \AA)       &  ~~~~4.38~~(3.487)~~~~   &  5.12~~(1.075)    \\[0.5ex]		
        X2C  & ~~~~vac                  &  ~~~~3.97~~(1.319)~~~~   &  4.52~~(1.740)    \\[0.5ex] 
		X2C  & ~~~~solv.~(6 \AA)        &  ~~~~4.45~~(2.685)~~~~   &  5.27~~(0.858)    \\[0.5ex]
		X2C  & ~~~~solv.~(6  \AA) + EEF &  ~~~~4.45~~(2.367)~~~~   &  5.27~~(0.785)    \\[0.5ex]
		X2C  & ~~~~solv.~(10 \AA)       &  ~~~~4.38~~(2.779)~~~~   &  5.27~~(0.883)    \\[0.5ex]
		X2C  & ~~~~solv.~(10 \AA) + EEF &  ~~~~4.38~~(2.412)~~~~   &  5.27~~(0.795)    \\[0.5ex]  
		\hline 
	\end{tabular}
\end{table}
Due to the rather dramatic effect of PE on the spectra, it  
is not meaningful to compare the solvent shift from the vacuum to the PE environment of the SR and X2C Hamiltonians. However, we can compare the final, solvated excitation energies and peak maxima: the excitation  energies (taken as the energy of the peak maxima) are not very different between SR and X2C Hamiltonians. 
However, the absorption cross sections at peak maxima  change between the Hamiltonians. For the 6 {\AA} system, the cross sections are 22\%-25\% lower for the X2C Hamiltonian.  Interestingly, the inclusion of a larger environment does not induce any significant changes, as seen when moving from the 6 {\AA}  to the 10 {\AA}  solvation sphere, with only the second peak being slightly shifted with 0.07 eV and the absorption cross-section increasing around 4 $\%$.
\begin{figure}[htb!]
\centering
\includegraphics[width=0.55\textwidth]{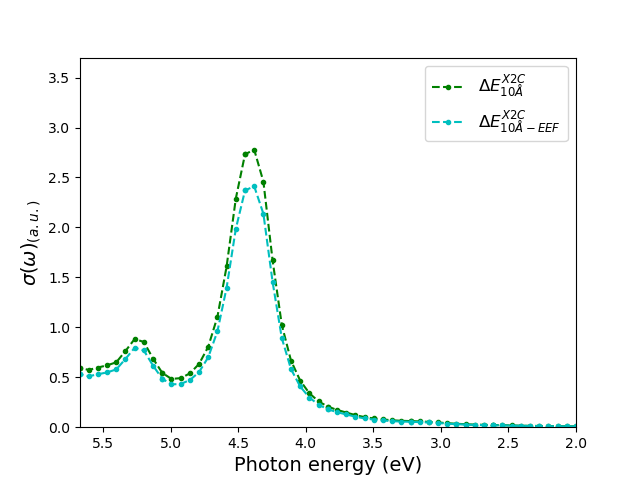}
 \caption{\footnotesize UV-Vis spectra calculated using CAM-B3LYP and X2C with and without EEF effects included. The total system was the 10 {\AA}, using the BP86 refined structure.}
\label{fig:eef_effects}
\end{figure}
We have also investigated how taking EEF effects into account changes the spectra (this investigation was only carried out for X2C). As expected, the excitation energies are not changed, while the absorption cross sections decrease with 9 and 13 $\%$ for the 10 {\AA}  solvation sphere (see Figure \ref{fig:eef_effects} and Table \ref{tab:solvated_static_struct_bp86}). For the 6 {\AA} solvation sphere, the corresponding changes in the absorption cross sections for the peaks are a decrease with 9 $\%$ and 11 $\%$ when including EEF effects (see Table \ref{tab:solvated_static_struct_bp86}). Thus, we conclude that we can safely employ a 6 {\AA}  sphere to study the solvent effect, but an EEF should be included. In the next subsection, we will additionally attempt to include a few solvent molecules in the QM region.

\subsection{Inclusion of water molecules in the QM region.}
\begin{figure}[htb!]
\centering
\begin{subfigure}[b]{0.5\textwidth}
\includegraphics[width=1.2\textwidth]{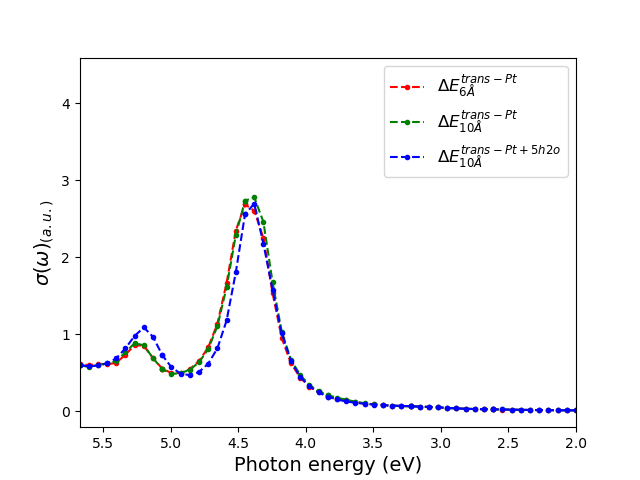}
\caption{\footnotesize}
\end{subfigure}
~~~~~\begin{subfigure}[b]{0.4\textwidth}
\centering
\includegraphics[width=0.8\textwidth]{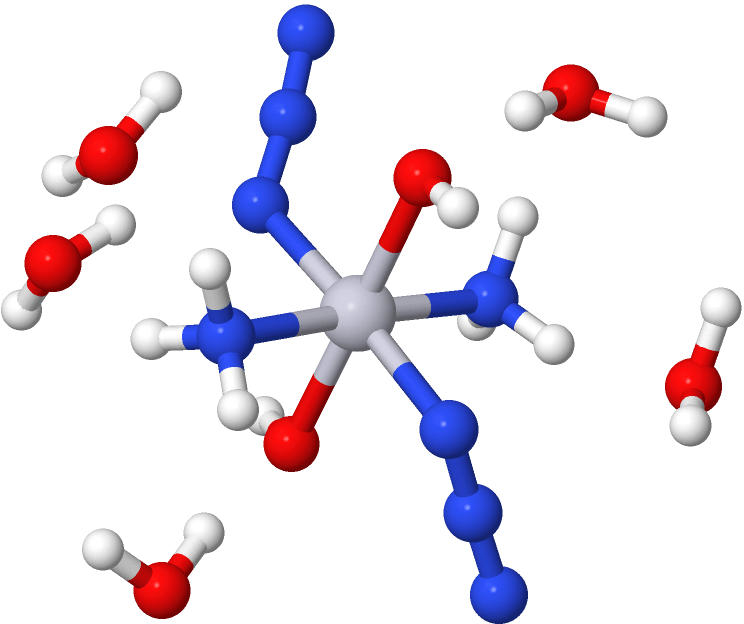}
\caption{\footnotesize}
\end{subfigure}
\caption{\footnotesize{(a) UV-vis absorption spectra calculated with a solvation sphere of 6 or 10 {\AA}  systems while employing PE and a QM region consisting of the \textit{trans}-Pt complex or the \textit{trans}-Pt and the closest 5 water molecules, shown in (b). All calculations were done X2C-CAM-B3LYP and with structures obtained from BP86 (in solvent). }}
\label{fig:water_in_qm}
\end{figure}

From the structure optimized with BP86, we included the five  water molecules closest to the \textit{trans}-Pt complex  in the QM region (all water molecules within a 4.2 Å distance from the Pt atom). The corresponding UV-vis spectra are shown in Figure \ref{fig:water_in_qm}(a) while the QM region is shown in \ref{fig:water_in_qm}(b). For the most intense peak, which is the only one seen experimentally, the effect of including QM water is much smaller than the effect of using a PBEh-3c refined structure. A slightly larger effect is seen for the higher energy peak, but the effect is still much smaller than the effect of using the refined structure. We carefully analyzed the response vectors to ensure that none of the excitations contained underlying transitions from the water molecules to the platinum complex (or vice versa). While this was not the case within the investigated frequency window, this will likely occur further into the high-energy region.  

\subsection{Dynamic effects: Conformations from MD}
\begin{figure}[htb!]
\hspace{-2cm}
 \begin{subfigure}[b]{0.48\textwidth}
\includegraphics[width=1.2\textwidth]{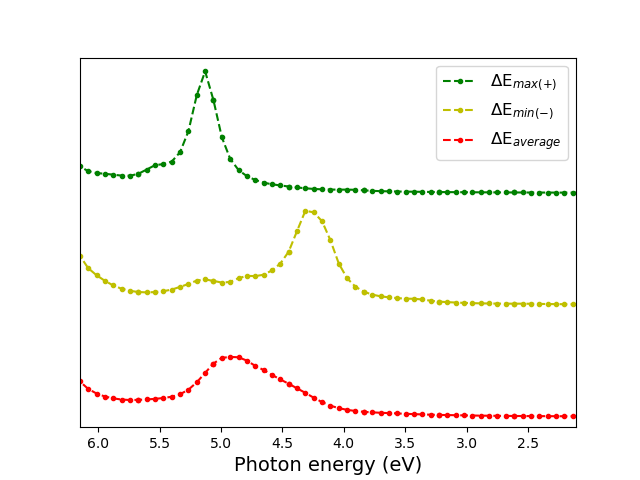}
\caption{\footnotesize}
\end{subfigure}
~~~~\begin{subfigure}[b]{0.48\textwidth}
\includegraphics[width=1.2\textwidth]{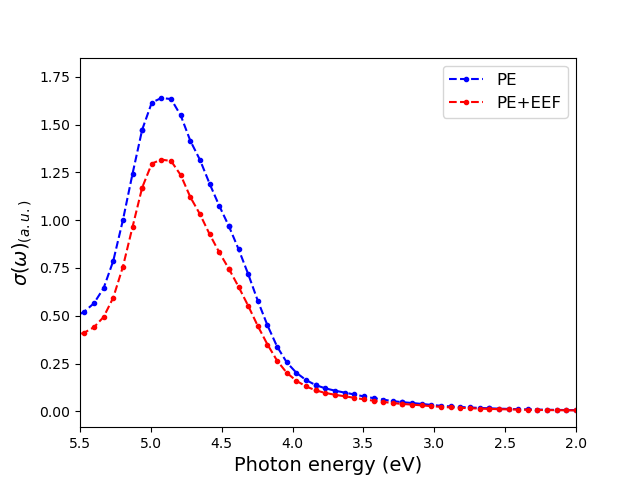}
\caption{\footnotesize}
\end{subfigure}
\caption{(a) Average spectra and the spectra for the structures that are shifted the most compared to the average peak. (b)  Comparison of the averaged spectra for PE with EEF and only PE.}
\label{fig:md_spect}
\end{figure}

We have thus far only investigated the solvent effects for a single conformation. While this did provide insight into how a solvent can influence the features of UV-Vis absorption spectra,  dynamic effects are required to mimic experimental conditions. From a series of snapshots obtained from a classical MD simulation with subsequent refinement through QM optimizations, we obtained 49 spectra with X2C. The spectra for the various snapshots can be seen in Figure S2; they generally have one peak (all peak positions are reported in Table S3). We here show only the mean spectrum, as well as the spectra with peak maxima most red- and blue-shifted from the mean, respectively (Figure \ref{fig:md_spect}a). The corresponding absorption cross sections and excitation energies at peak maxima are provided in Table \ref{tab:peak_data}. From the spectra that display the largest difference in the position of the peak maxima, the span of potential peak maxima ranges from 4.31 to 5.13 eV, i.e., 0.8 eV. This demonstrates the importance of dynamic solvent effects.\\ 
All spectra calculations were carried out both with and without EEF effects. We mainly consider the values including the EEF effects, but note that the absorption cross-section decrease with 24 $\%$ when we include the EEF effect. The mean spectra with and without EEF are shown in Figure \ref{fig:md_spect}(b). 

Previous investigations employing range-separated functionals found that these calculations provide excellent agreement with experimental results for both the shape of the spectra and the energy at the absorption maximum\cite{sokolov2011,creutzberg2020}. However, these investigations were carried out without inclusion of dynamical and electronic solvent effect. In light of our present results, it is clear that  benchmarking should include these effects. Yet, the inclusion of dynamics does not bring the computed spectra closer to the peak maximum of the experimental spectrum. The computed mean peak position is 
4.79  $\pm$ 0.23 eV, compared to the experimental 4.35 eV.\cite{mackay2006} Several underlying causes for this apparent discrepancy may be found: one of them being the choice of functional. To investigate this possibility, we computed the spectra for 5 structures using B3LYP, employing  5 randomly selected snapshots (see Figure S4). We find that the peaks of the CAM-B3LYP spectra are blue-shifted compared to those of B3LYP. The average blue-shift is calculated to be 0.27 eV. If the calculated average of 4.79 eV from Table \ref{tab:peak_data} is shifted with 0.27 eV, the calculated result is within the experimental results (including the uncertainty). Thus, the discrepancy is likely caused at least partially by a tendency for CAM-B3LYP to blue shift excitation energies, and this blue shift likely cancels with the lack of dynamical and electronic solvent effects in previous studies.\cite{sokolov2011,creutzberg2020} Moreover, a blueshift of CAM-B3LYP has been observed previously: A benchmark study of vertical excitation energies for TD-DFT against CC2 found that several functionals (among them CAM-B3LYP) overestimated vertical excitation energies with 0.2-0.3 eV \cite{shao2020}. Another benchmark study (focusing on transition metal complexes) yielded a similar conclusion\cite{latouche2015}.  For three of the five snapshots used to compare B3LYP and CAM-B3LYP, we also investigated the effect of extending the basis set to triple-zeta quality, but this cannot explain the blue shift (see Figure S5). 

We also note that some of  the deviation from experiment in the present investigation may be caused by inaccuracies in the sampled structures from the subsequent restricted QM refinement.  We investigated if the blueshift could be correlated with a simple structural parameter. In the supporting information (Figure S6), we show how the peak maximum depends on the \ce{Pt-N3} bond distance for the considered snapshots. This distance was chosen due to LMCT nature of the transition and the involvement of orbitals located on the azide ligands. 
The peak maxima are indeed sensitive to this distance, and distances smaller than the average usually result in  transitions energies above the average. However, since we do not have a reference structure, it is not possible to isolate the blueshift to be an issue caused by the structure alone. We therefore leave this matter here.  A reference structure may be obtained from \textit{ab initio} MD, although it will come at the price of reduced sampling time or significantly increased computational demands. 
Despite the blueshift compared to experiment, our present results serve to quantify the importance of environment interactions on calculations of molecular properties, and therefore further emphasize the need for an explicit treatment of the environment.  
\begin{table}[htb!]
	\centering
	\caption{\footnotesize The mean maximum peak position displayed together with the peak positions of the average spectra and the spectra for the structures that deviate the most from the mean maximum peak position (seen in Figure \ref{fig:md_spect}b). All reported values include EEF effects. } 
	\label{tab:peak_data}
	\begin{tabular}{lccc}
		\hline
		\hline \\[-2.0ex]
		Peak~~~~ & ~~~~Energy [eV]~~~~  & ~~~~$\sigma (\omega)$ [a.u]~~~~  \\ [0.5ex] 
		\hline \\[-1.5ex]
		 Mean max.  & 4.79  $\pm$ 0.23 & 2.33 $\pm$ 0.52\\[0.5ex]
	     $\Delta E_{max(+)}$   & 5.13 & 2.72 \\[0.5ex]
		 $\Delta E_{max(-)}$ & 4.31 & 2.10 \\[0.5ex]
		\hline
	\end{tabular}
\end{table}

Finally, we discuss whether our results are converged with respect to the sampling of solvent configurations. We have investigated the convergence in Table S2, where the average peak maxima and excitation energies are compiled in blocks of 10, 20, 30 (and so forth) snapshots. The calculated mean excitation energies and absorption cross sections are converged after 20 snapshots. Thus, we do not expect significant errors due to the employed sampling method.

\section{Conclusion}
In this study we have investigated UV-vis absorption spectra of the \textit{trans}-\textit{trans}-\textit{trans}-[\ce{Pt(N3)2(OH)2(NH3)2]} (\textit{trans}-Pt) complex embedded in various aqueous environments,  using the relativistic CPP method combined with the PE model. The first part of the study involves a single structure based on \textit{trans}-Pt optimized in water, employing either  PBEh-3c or BP86 to refine the structure. The underlying structures have a large effect on the resulting spectra, and we chose here the theoretically best method (BP86). Thus,  the following analyzes employ structures obtained at the BP86 level. We also find that the spectrum computed without electronic solvent effects is rather different, compared to those including PE. However, extending the environment from 6 {\AA} to 10 {\AA}  only has a minor effect on the spectra, leading to the conclusion that the smaller solvation sphere of 6 {\AA}  is sufficient. Extending the QM region by including the nearest five water molecules likewise has only a benign effect on the calculated spectrum. 

Regarding the inclusion of dynamics (with subsequent QM optimizations), we find that the individual snapshots lead to rather different spectra. While this emphasizes that explicit treatment of the environment is required, the resulting average excitation energy (taken as the average peak maxima positions) is somewhat blue-shifted, compared to the experiment. Yet, the excellent agreement with experiment in previous investigations with range-separated functionals\cite{sokolov2011,creutzberg2020} is in light of our present results likely to be somewhat fortuitous due to the lack of dynamic (and electronic) solvent effects.  Indeed, employing global hybrid functionals such as B3LYP yield spectra that are systematically red-shifted compared to CAM-B3LYP. This shows the importance of including a realistic solvent environment in functional benchmarking. Finally, EEF effects significantly alter the absorption cross sections and should be included.

\section*{Conflicts of interest}
There are no conflicts of interest to declare.

\section*{Acknowledgments}

EDH thanks The Villum Foundation, Young Investigator Program (grant no. 29412), the Swedish Research Council (grant no. 2019-04205), and Independent Research Fund Denmark (grants no. 0252-00002B and no. 2064-00002B) for support. 

\bibliographystyle{vancouver}
\bibliography{article}

\newpage

\end{document}



\title[]{Supporting information: A method to capture the large relativistic and solvent effects on UV-vis spectra of photo-activated metal complexes}

\author{Joel Creutzberg}
\email{joel.creutzberg@teokem.lu.se}
\affiliation{Division of Theoretical Chemistry, Lund University, Lund, Sweden}
\author{Erik Donovan Hedeg{\aa}rd$^{1,}$}
\email{erdh@sdu.dk}
\affiliation{Department of Physics, Chemistry and Pharmacy, Campusvej 55, 5230 Odense, Denmark}

\date{\today}

\maketitle

\begin{figure}[H]
 \begin{subfigure}[b]{0.5\textwidth}
\includegraphics[width=1\textwidth]{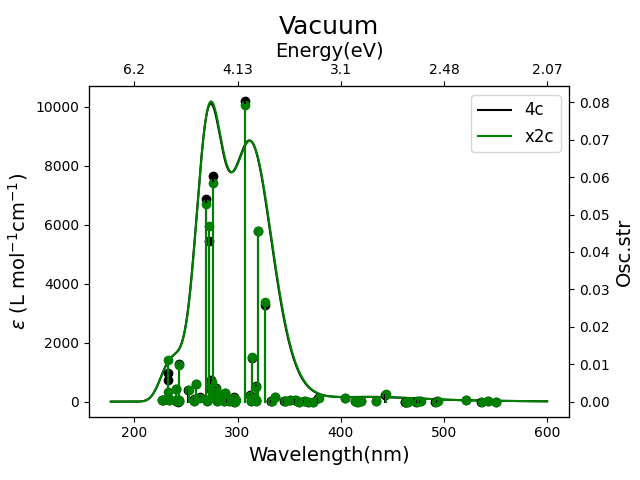}
\caption{\footnotesize}
\end{subfigure}
 \begin{subfigure}[b]{0.5\textwidth}
\includegraphics[width=1\textwidth]{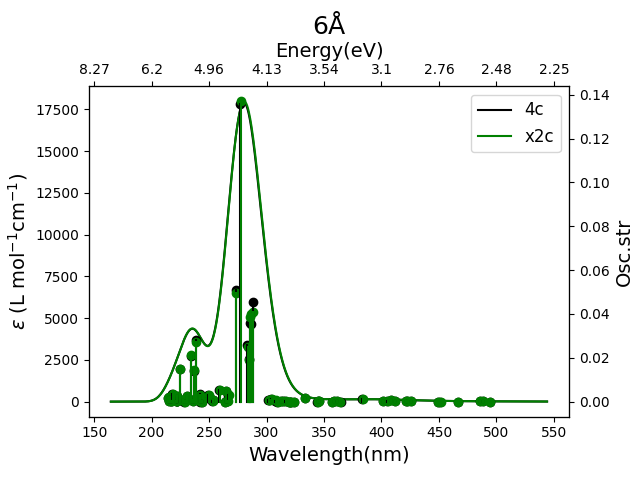}
\caption{\footnotesize }
\end{subfigure}
 \begin{subfigure}[b]{0.5\textwidth}
\includegraphics[width=1\textwidth]{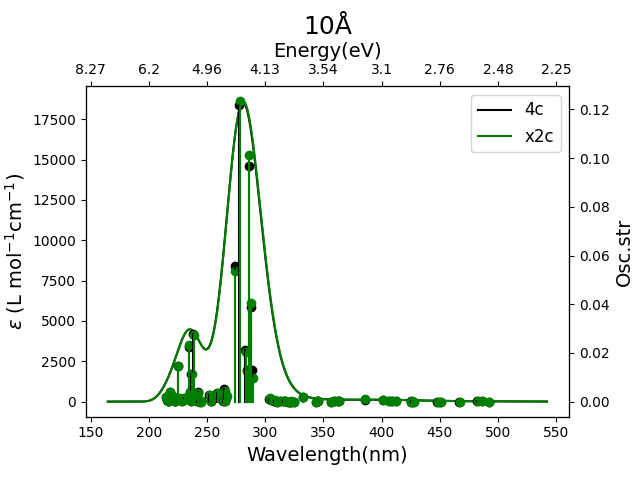}
\caption{\footnotesize }
\end{subfigure}

 \caption*{Figure S1: \footnotesize UV-vis spectra calculated using TDDFT with the CAM-B3LYP functional comparing X2C with 4C for (a) Complex in vacuum taken from structure optimized with BP86 (b) Complex solvated in a 6 Å sphere of water molecules (optimized with BP86) (c) Complex solvated in a 10 Å sphere of water molecules (optimized with BP86). }
\label{fig:solvent_effects_x2c_4c}
\end{figure}

\begin{figure}[H]
\includegraphics[width=0.8\textwidth]{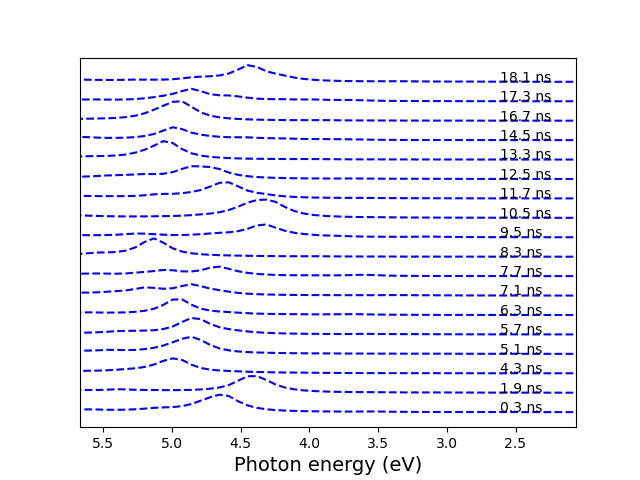}
\caption*{Figure S2: \footnotesize Spectra calculated for various snapshots from md simulation. Peak positions are reported in Table S1 }
\end{figure}

\begin{table}[H]
\centering
\scalebox{0.8}{
\begin{tabular}[t]{ c c c}
\hline
\hline
Snapshot ~~~ & Peak energy (eV) ~~~ & $\sigma (\omega)$ [a.u] \\
\hline
0.3 ns & 4.653 & 2.586 \\ [-1.5ex]
1.3 ns & 4.517 & 3.256 \\ [-1.5ex]
1.7 ns & 4.517 & 2.363 \\ [-1.5ex]
1.9 ns & 4.381 & 2.458 \\ [-1.5ex]
2.9 ns & 4.857 & 2.425 \\ [-1.5ex]
3.9 ns & 4.313 & 2.099 \\ [-1.5ex]
4.3 ns & 4.993 & 2.21 \\ [-1.5ex]
4.5 ns & 4.993 & 2.911 \\ [-1.5ex]
4.9 ns & 4.585 & 1.347 \\ [-1.5ex]
5.1 ns & 4.857 & 2.531 \\ [-1.5ex]
5.3 ns & 4.789 & 1.537 \\ [-1.5ex]
5.7 ns & 4.857 & 2.437 \\ [-1.5ex]
5.9 ns & 4.653 & 3.079 \\ [-1.5ex]
6.1 ns & 4.993 & 2.571 \\ [-1.5ex]
6.3 ns & 4.925 & 2.319 \\ [-1.5ex]
6.5 ns & 4.789 & 1.857 \\ [-1.5ex]
6.7 ns & 4.993 & 2.527 \\ [-1.5ex]
7.1 ns & 4.857 & 1.689 \\ [-1.5ex]
7.3 ns & 4.993 & 1.357 \\ [-1.5ex]
7.5 ns & 4.789 & 2.182 \\ [-1.5ex]
7.7 ns & 4.653 & 1.414 \\ [-1.5ex]
7.9 ns & 4.653 & 1.732 \\ [-1.5ex]
8.1 ns & 4.517 & 2.533 \\ [-1.5ex]
8.3 ns & 5.129 & 2.716 \\ [-1.5ex]
8.7 ns & 5.061 & 2.84 \\ [-1.5ex]
9.5 ns & 4.313 & 1.908 \\ [-1.5ex]
10.3 ns & 4.993 & 2.412 \\ [-1.5ex]
10.5 ns & 4.313 & 2.704 \\ [-1.5ex]
10.9 ns & 5.129 & 3.029 \\ [-1.5ex]
11.3 ns & 4.653 & 1.403 \\ [-1.5ex]
11.7 ns & 4.585 & 2.383 \\ [-1.5ex]
11.9 ns & 5.129 & 2.839 \\ [-1.5ex]
12.1 ns & 4.857 & 2.786 \\ [-1.5ex]
12.5 ns & 4.857 & 1.894 \\ [-1.5ex]
12.9 ns & 5.061 & 1.287 \\ [-1.5ex]
13.1 ns & 5.061 & 2.64 \\ [-1.5ex]
13.3 ns & 5.061 & 2.725 \\ [-1.5ex]
13.5 ns & 4.789 & 2.131 \\ [-1.5ex]
14.5 ns & 4.993 & 1.903 \\ [-1.5ex]
15.1 ns & 4.857 & 2.215 \\ [-1.5ex]
16.1 ns & 4.653 & 2.879 \\ [-1.5ex]
16.7 ns & 4.925 & 2.874 \\ [-1.5ex]
16.9 ns & 4.789 & 2.952 \\ [-1.5ex]
17.1 ns & 4.381 & 2.6 \\ [-1.5ex]
17.3 ns & 4.857 & 1.846 \\ [-1.5ex]
17.7 ns & 4.789 & 2.775 \\ [-1.5ex]
17.9 ns & 4.857 & 2.902 \\ [-1.5ex]
18.1 ns & 4.449 & 2.458 \\ [-1.5ex]
18.3 ns & 5.061 & 1.615 \\ [-1.5ex]
\end{tabular}
}
\caption*{Table S1: Data for the peaks of all md snapshots }
\end{table}

\begin{figure}[H]
 \begin{subfigure}[b]{0.5\textwidth}
\includegraphics[width=1\textwidth]{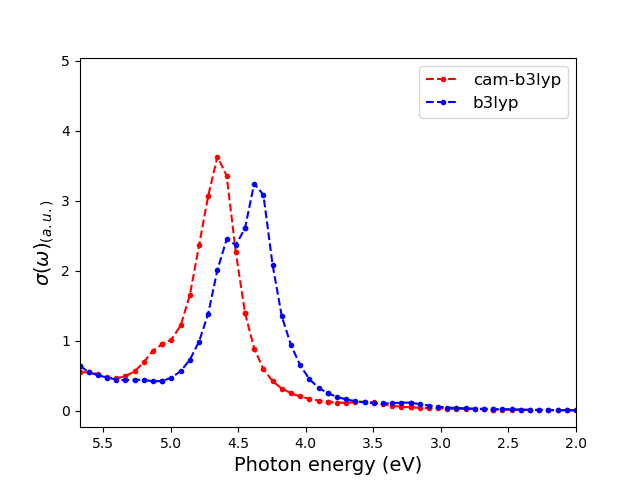}
\caption{\footnotesize }
\end{subfigure}
 \begin{subfigure}[b]{0.5\textwidth}
\includegraphics[width=1\textwidth]{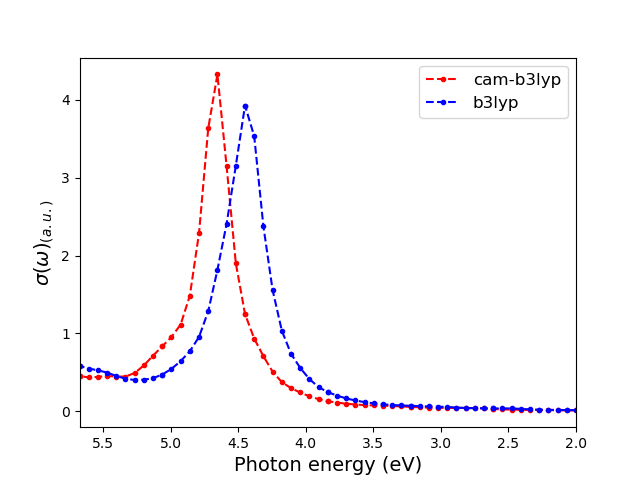}
\caption{\footnotesize }
\end{subfigure}
 \begin{subfigure}[b]{0.5\textwidth}
\includegraphics[width=1\textwidth]{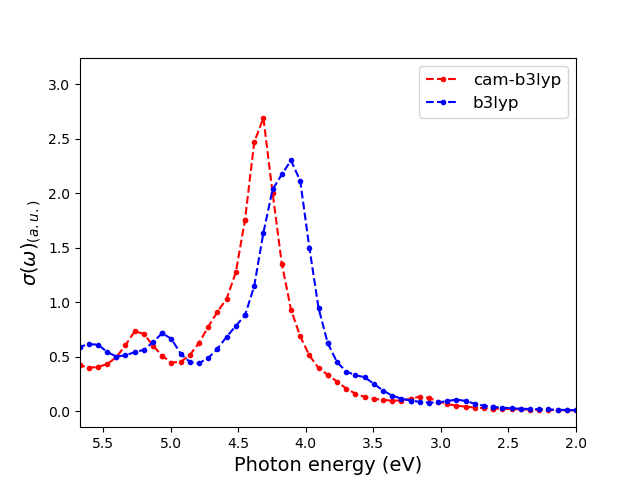}
\caption{\footnotesize }
\end{subfigure}
 \begin{subfigure}[b]{0.5\textwidth}
\includegraphics[width=1\textwidth]{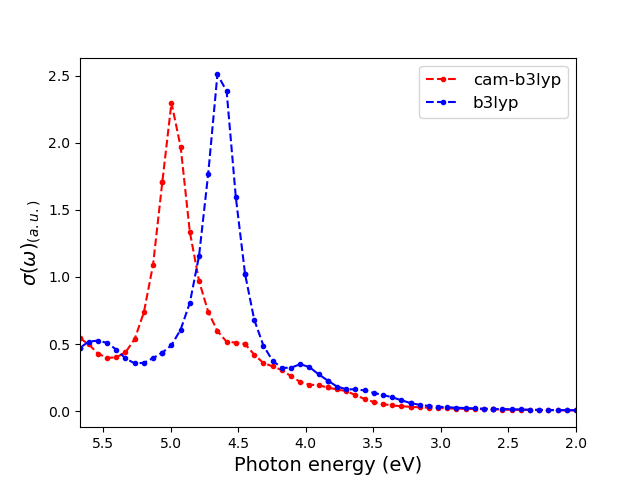}
\caption{\footnotesize }
\end{subfigure}
 \begin{subfigure}[b]{0.5\textwidth}
\includegraphics[width=1\textwidth]{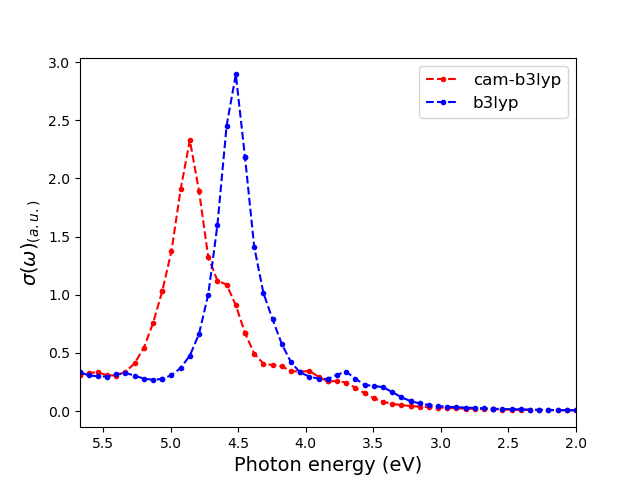}
\caption{\footnotesize }
\end{subfigure}
 \caption*{Figure S4 \footnotesize Comparison of B3LYP and CAM-B3LYP for 5 structures  }
\end{figure}

\begin{figure}[H]
 \begin{subfigure}[b]{0.5\textwidth}
\includegraphics[width=1\textwidth]{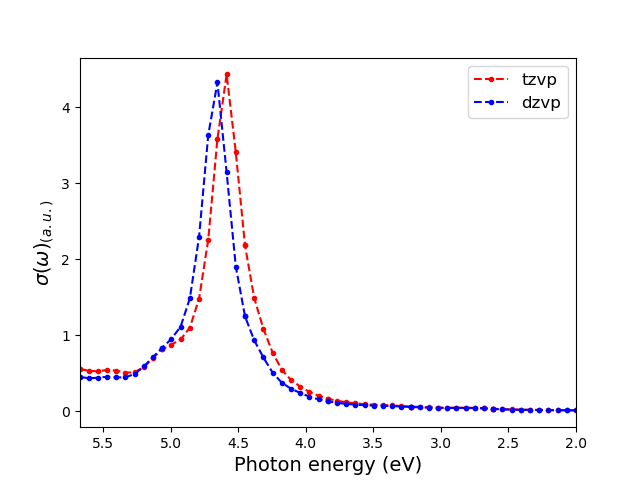}
\caption{\footnotesize }
\end{subfigure}
 \begin{subfigure}[b]{0.5\textwidth}
\includegraphics[width=1\textwidth]{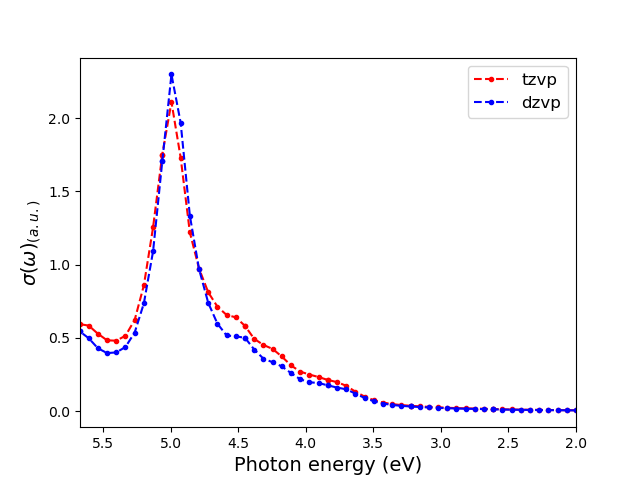}
\caption{\footnotesize }
\end{subfigure}
 \begin{subfigure}[b]{0.5\textwidth}
\includegraphics[width=1\textwidth]{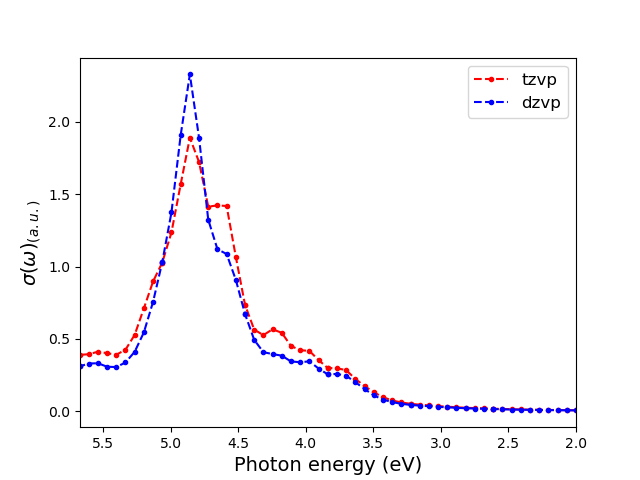}
\caption{\footnotesize }
\end{subfigure}
 \caption*{Figure S5 \footnotesize Comparison of spectra obtained with DZVP (def2-sv(p) for ligands and dyall.v2z for Pt) and TZVP (def2-tzvp for the ligands and dyall.v3z for Pt)  using CAM-B3LYP for 3 structures.  }
\end{figure}

\begin{figure}[H]
 \begin{subfigure}[b]{0.5\textwidth}
\includegraphics[width=1\textwidth]{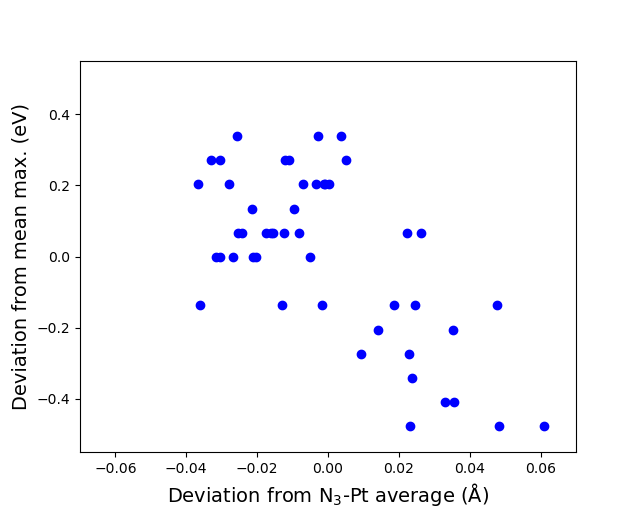}
\caption{\footnotesize }
\end{subfigure}
 \begin{subfigure}[b]{0.5\textwidth}
\includegraphics[width=1\textwidth]{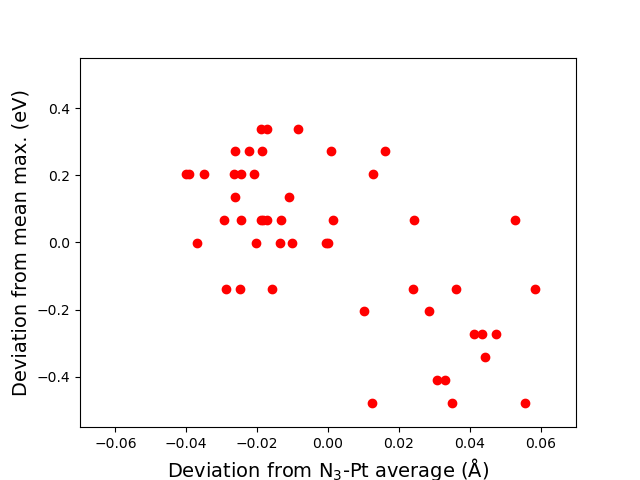}
\caption{\footnotesize }
\end{subfigure}

 \caption*{Figure S6 \footnotesize The figures show the deviation from the average peak displayed against the deviation from the average bond length with respect to the two N$_3$-Pt bond distances. }
\end{figure}

\begin{table}[H]
\centering
\begin{tabular}[t]{c c c | c c c}
\hline
\hline
Snapshots & ~~~~Peak (eV)~~~~& ~~~~$\sigma(\omega)$~~[a.u.]~~~~& ~~~~Snapshots~~~~ &  Peak (eV)~~~~& ~~~~$\sigma(\omega)$~~[a.u.]  \\ [1.5ex]
\hline
1--10  & 4.67 $\pm$ 0.23  & 2.42 $\pm$ 0.48  & 1--10 & 4.67 $\pm$ 0.23 & 2.42 $\pm$  0.48 \\ [2.5ex]
11--20 & 4.87 $\pm$ 0.11  & 2.15 $\pm$ 0.54  & 1--20 & 4.76 $\pm$ 0.21 & 2.29 $\pm$  0.52 \\ [2.5ex]
21--30 & 4.76 $\pm$ 0.27  & 2.22 $\pm$ 0.47  & 1--30 & 4.76 $\pm$ 0.23 & 2.27 $\pm$  0.51 \\ [2.5ex]
31--40 & 4.87 $\pm$ 0.26  & 2.37 $\pm$ 0.59  & 1--40 & 4.79 $\pm$ 0.24 & 2.30 $\pm$  0.53 \\ [2.5ex]
41--49 & 4.78 $\pm$ 0.17  & 2.46 $\pm$ 0.42  & 1--49 & 4.79 $\pm$ 0.23 & 2.33 $\pm$  0.52 \\ [2.5ex]
\hline"
\end{tabular}
\caption*{TABLE S2: Averages and deviations for excitation energies and absorption cross sections (either blocks of 10 snapshots  or cumulative).}
\end{table}